# From Clicks to Conversions: Analysis of Traffic Sources in E-Commerce

Amrutha Muralidhar*

Dept. of Computer Science and Engineering

B.M.S. College of Engineering

Bengaluru, India

Yathindra Lakkanna

Dept. of Fashion and Lifestyle Accessory Design

National Institute of Fashion Technology

Bengaluru, India

## Abstract

Over the past years, e-commerce platforms have expanded substantially, providing customers with convenient shopping experiences. To enhance e-commerce websites, it's essential to grasp user engagement and factors affecting conversion rates. This optimization is achieved by aligning platforms with user expectations, thereby fostering successful online shopping experiences, and contributing to sustained growth in the dynamic digital marketplace. In this paper, we conduct a comprehensive analysis focusing on user interactions, conversion metrics, and the entire user journey within an e-commerce platform. Our exploration spans exit rates and sessions across different devices and browsers, conversion rates for various traffic sources, and the user journey from product details to checkout. Findings suggest a need for targeted improvements in mobile optimization and browser compatibility, indicated by higher exit rates on mobile devices and varying rates on different browsers. The conversion rate analysis emphasizes the varying effectiveness of traffic sources, highlighting the potential of advertised mediums, while identifying areas for improvement in referral and affiliate traffic. Examining the user journey showed potential bottlenecks in the conversion process, the identified gap was between user interest and completed transactions. Our study suggests improving the checkout process strategically to streamline user transactions. These insights provide actionable guidance for businesses seeking to refine their platforms and optimize performance in the ever-evolving landscape of e-commerce.

**Keywords:** *User Behavior Analysis, Transaction Optimization, Conversion Rates, Funnel Analysis, E-Commerce*

* Corresponding author

## 1. Introduction

The digital landscape has transformed the way commerce operates, with an increasing reliance on online platforms for buying and selling goods and services [1]. As e-commerce continues to flourish, businesses are confronted with the challenge of not only attracting a substantial volume of traffic to their websites but, more importantly, converting that traffic into meaningful transactions [2]. The journey from clicks to conversions represents a critical juncture in the success of any e-commerce venture.

Understanding the diverse sources that drive traffic to e-commerce websites is paramount for businesses seeking to optimize their online presence [3]. From organic search results and paid advertisements to social media referrals and direct visits, each traffic source carries its unique set of implications for conversion rates. This paper delves into the intricate relationship between these traffic sources and the goal of e-commerce: the conversion of visitors into customers.

The analysis of traffic sources in e-commerce is not merely an exercise in data interpretation; it is a strategic imperative. As businesses navigate a dynamic and competitive online marketplace, decoding the patterns and behaviors associated with different traffic sources becomes essential for informed decision-making [4]. This paper aims to contribute to the existing body of knowledge by providing a comprehensive examination of the impact of various traffic sources on conversion rates, offering insights that can guide e-commerce businesses in crafting effective strategies.

In the pages that follow, we will explore the existing literature on e-commerce traffic sources, outline our methodology for data collection and analysis, categorize and describe different types of traffic sources, and present findings that shed light on the factors influencing conversions. By the end of this exploration, we aspire to equip businesses with actionable recommendations for optimizing their traffic sources and, consequently, enhancing their conversion rates in the dynamic world of e-commerce.

## 2. Literature Review

In the age of digital media, the management of customer journeys in e-commerce has become a complex task due to the vast digital information sources accessible to consumers. Klein et al. (2020) [5] address the challenge of quantifying customers' cross-media exposure and its impact on individual customer journeys. They introduce "media entropy" as a metric, drawing on information and signaling theory, to measure the entropy of company-controlled and peer-driven media sources. Through a probit model analysis, they demonstrate that cross-media exposure significantly influences purchase decisions, especially in digital environments, for customers not currently owning the brand and for perceived weaker brands.

Kakalejčík, Bucko, and Danko (2020) [6] contribute to the understanding of online customer journeys by investigating the impact of newly created brand awareness on buying behavior. Analyzing over 280,000 online customer journeys from Slovakian e-commerce stores, they introduce the concept of "direct traffic effect" based on theoretical brand awareness criteria. Their findings reveal that these criteria significantly contribute to the company's revenues, emphasizing the importance of positive customer experiences in the early stages of interactions online.

Duan and Zhang (2021) [7] focus on the evaluation of the effectiveness and return on investment (ROI) across different referral channels in e-commerce. Employing the vector autoregressive (VARX) model on a large-scale clickstream dataset, they explore the dynamics and interdependencies among search engines, social media, and third-party websites. Their findings highlight the immediate and cumulative effects of social media referrals on conversion rates, providing valuable insights for digital marketing managers assessing the economic value of online referral channels.

Cao, Chu, Hui, and Xu (2021) [8] contribute to the understanding of online marketing strategies by examining the relationship between referral marketing and price promotion. Their research on a large e-commerce platform reveals that although referral marketing increases sales, its effectiveness is attenuated by price promotion. The study emphasizes the need for marketing managers to scrutinize the complementarity of promotional strategies, especially in addressing consumers' quality concerns in the online environment.

Zumstein and Kotowski (2020) [9] provide practical insights into the success factors of e-commerce by offering ten recommendations for online shops. These recommendations range from offering free shipping and personalized content to analyzing digital analytics data. The study emphasizes the importance of building trust, personalizing user experiences, and leveraging various sales and marketing channels for effective e-commerce operations.

Azoev and Khokhlov (2022) [10] delve into the intricacies of site-based conversion, with a particular focus on identifying criteria for the fundamental conversion of a website, independent of advertising campaign elements. The study addresses challenges related to the accuracy of media plans and proposes a classification of advertising traffic sources. By calculating conversion rejection coefficients for different advertising formats, the authors aim to empower media planners to develop more accurate plans, minimizing the need for experimental campaigns. This contribution sheds light on the factors influencing conversion rates and sets the stage for further research to enhance the precision of media plans in the digital landscape.

Spais (2010) [11] explores the extension of Bedny's [12] perspective of 'activity' theory as a framework for understanding and analyzing new online promotion channels, specifically search engines. The conceptualization of activity theory is applied to Search Engine Optimization (SEO) contextual issues, providing insights into the design and analysis of SEO promotion techniques. The findings contribute to the literature by offering practical implications for promotion managers seeking to navigate the dynamic landscape of SEO as an online promotion technique.

Kang et al. (2021) [13] delve into the unexplored dynamics of live streaming commerce, emphasizing interactivity's impact on customer engagement. The study, guided by the stimulus-organism-response (S-O-R) framework, identifies a curvilinear relationship between interactivity and customer engagement. Tie strength emerges as a key mediator, showcasing the nuanced role of user interactions in the context of live streaming commerce platforms.

Rosário and Raimundo (2021) [14] underscore the transformative influence of technological developments. The study explores how user-friendly design features on e-commerce websites enhance interactivity and engagement. The integration of social networking sites (SNSs) fosters two-way communication, enabling information sharing and innovation. Moreover, the integration of IT and big data technologies enables personalization and customization, ensuring the adaptability of e-commerce businesses in a competitive global landscape.

Sakas and Giannakopoulos (2021) [15] delve into the significance of mobile and desktop device usage in shaping the digital brand name of airline firms. The research underscores the impact of these devices on organic, referral, and direct traffic, providing insights into the nuanced effects of mobile and desktop usage on airlines' digital brand name.

Dolega et al. (2021) [16] contribute to the understanding of social media's impact on business outcomes using a major online retailer's data. Findings highlight that while social media increases web traffic, its influence on product orders and sales is limited. Larger campaigns, especially on Facebook, yield higher orders and sales, emphasizing the nuanced effectiveness of social media marketing that varies based on campaign size and product characteristics.

Hasan et al. (2009) [17] identified thirteen web metrics as indicators of potential usability problems on e-commerce websites, providing a quick, easy, and cost-effective overview. While these metrics offer continual insights into a site's usability, the authors emphasized the need for additional techniques, such as heuristic evaluation, for a thorough understanding of specific issues. They proposed a future research framework incorporating Google Analytics (GA) to identify usability problems swiftly and inexpensively, complemented by heuristic evaluation for detailed analysis.

These studies collectively contribute to the understanding of conversion in e-commerce, exploring the impact of advertising traffic sources on conversion rates and presenting an innovative application of activity theory to the realm of Search Engine Optimization. As the digital marketing landscape evolves, these insights become crucial for practitioners and researchers alike, offering valuable perspectives for the development of effective promotion strategies in the online domain.

## 3. Methodology

### 3.1. Research Design

This research adopts a retrospective study design, leveraging data extracted from the Google Analytics platform. This approach allows for a systematic examination of historical data, enabling a detailed analysis of user behavior patterns over time. By focusing on past user interactions, the study aims to uncover insights into the website's performance and user engagement on the e-commerce platform.

### 3.2. Data Source

Our study relies on Google Analytics as the primary data source, a web analytics tool known for its capability to capture detailed metrics associated with user interactions. This encompasses critical metrics, including landing page interactions, bounce rates, exit rates, and conversion data. Google Analytics was chosen for its robust features, providing a comprehensive dataset essential for the in-depth analysis of user behavior on the e-commerce website. The platform's capacity to capture nuanced user behavior makes it an ideal source for this investigation.

### 3.3. Duration of the Study

Data used spans a two-year period, providing a longitudinal perspective for evaluating trends, patterns, and user behaviors. The extended duration allows for a thorough examination of potential seasonality and long-term dynamics in user engagement.

### 3.4. Key Metrics Definitions

- **Bounce Rate**

The bounce rate, a crucial metric, is computed as the number of sessions where users entered the site and immediately left divided by the total number of sessions [18]. A high bounce rate may signal issues related to user expectations, SEO, or landing page functionality. Analyzing this metric aids in pinpointing potential areas of improvement.

- **Exit Rate**

The exit rate, calculated as the number of sessions where users navigated to a page and then exited the site divided by the total number of sessions, provides insights into user interactions with specific pages. Elevated exit rates may indicate issues with page content, functionality, or overall user experience, guiding optimization efforts.

- **Conversion Rate**

The conversion rate, a pivotal metric, represents the percentage of sessions resulting in a completed transaction. This metric is fundamental for evaluating the effectiveness of the website's conversion process. Understanding conversion rates allows for targeted enhancements to optimize the user journey.

### 3.5. Query Execution

Customized queries were formulated to extract pertinent data from the Google Analytics dataset. These queries specifically targeted elements such as traffic mediums, browsers, devices, and essential metrics like sessions, exit rates, transactions, and conversion rates.

### 3.6. Analysis Techniques

- **Descriptive Analysis**

A descriptive analysis was conducted on the extracted data to provide a comprehensive overview of key metrics. This step aimed to establish a foundational understanding of the e-commerce website's overall performance.

- **Comparative Analysis**

Comparative analysis explored variations in bounce rates, exit rates, and conversion rates across different dimensions such as landing pages, browsers, devices, and traffic sources, providing actionable insights for improvement.

- **Conversion Path Analysis**

Tracing the customer's conversion path involved a detailed examination of user behavior from landing pages to checkout. This analysis included scrutinizing the progression of users through product details, checkout, and purchase completion.

Identifying potential bottlenecks in the conversion process is helpful for strategic optimizations.

- **User Engagement Analysis Across Devices**

  Analysis was conducted to uncover specific issues related to user engagement across various devices, including tablets, desktops, and mobiles. This qualitative approach aimed to identify issues at the browser, device, page, form, and button levels, offering nuanced insights into the user experience and guiding responsive design improvements.

### 3.7. Data Visualization

The findings from the analysis were effectively visualized using plots and charts. This visualization facilitated a clear and interpretable representation of insights gained from the data, aiding in the communication of key findings.

## 4. Findings

Figure 1 displays the bounce rates for the website's top 10 pages, while Figure 2 illustrates their exit rates. The "Men's Apparel" page exhibits a lower bounce rate, suggesting users engage with its content. However, Figure 2 reveals a higher exit rate for the same page. This indicates that, despite initial engagement, a significant portion of visitors exit the website after interacting with the page. The discrepancy between bounce and exit rates signals potential challenges in the user journey or conversion process, warranting further investigation for optimization.

Figure 3 presents an overview of exit rates and sessions categorized by different devices. The analysis reveals that mobile devices exhibit an exit rate notably higher than other devices. This suggests a potential optimization opportunity for mobile viewing. Most sessions are from desktop users, indicating the need to explore potential reasons for the higher exit rate on mobile.

Figure 4 illustrates the distribution of sessions and exit rates across various browsers. Chrome dominates in sessions, with a comparatively lower exit rate, while Safari and Firefox exhibit higher exit rates.

The conversion rate is a crucial indicator for businesses to understand the effectiveness of their marketing efforts and the user experience on their platforms. It provides insights into the percentage of sessions that result in completed transactions, which is a key factor in determining the success of an online business [19]. Figure 5 presents the conversion rates across different traffic mediums, highlighting the varying performance of each medium in terms of converting user sessions into completed transactions. The conversion rate (CR) formula is expressed as:

$$CR = \frac{Number\ of\ Sessions\ with\ Transactions}{Total\ Number\ of\ Sessions} \times 100$$

The data indicates that the CPM (Cost Per Mille) medium, followed by CPC (Cost Per Click), demonstrates higher conversion rates. This suggests that users coming from these traffic sources are more likely to complete transactions compared to other mediums. On the other hand, referral and affiliate traffic exhibit lower conversion rates, indicating potential areas for improvement in engaging and converting users from these sources.

Figure 6 illustrates the conversion path, showing the progression of users through various actions leading up to a completed purchase. It is notable that a significant number of users proceed from product detail views to the checkout stage, indicating active interest in making a purchase. However, only a fraction of these users completes the purchase, pointing towards potential barriers or challenges in the final stages of the conversion process.

The most common user action observed is "Click through of product lists," indicating active engagement with the products offered. On the other hand, "Completed purchase" is identified as the least common action, suggesting that there may be obstacles hindering users from finalizing their purchases.

This analysis of conversion rates and user behavior provides valuable insights for businesses to optimize their marketing strategies and user experience. By understanding which traffic mediums perform better in terms of conversions and identifying potential bottlenecks in the conversion path, businesses can make informed decisions to enhance their overall conversion rates and ultimately drive more successful transactions.

## 5. Discussion

Based on the data provided in Figures 3 and 4, there are several optimization opportunities to improve user experience and conversion rates. For mobile devices, which have a higher exit rate, it is essential to identify the reasons for this trend. Possible factors could include slow loading times, poor user interface design, or lack of mobile optimization. Addressing these issues could lead to a decrease in the mobile exit rate and an increase in overall user engagement. For browsers, the higher exit rates on Safari and Firefox compared to Chrome suggest that there might be compatibility issues or performance differences between these browsers and the website. Investigating these discrepancies and implementing necessary changes can help improve user experience and reduce exit rates across all browsers. The observed progression rate from product details to the checkout page highlights the importance of product presentation and customer engagement strategies employed by e-commerce platforms. However, the relatively lower conversion rate from the checkout page to completed purchases raises questions about the platform's ability to effectively convert user interest into actual transactions.

This study's findings indicate a potential disconnect between the initial interest in product details and the ultimate commitment to making a transaction. It is crucial for e-commerce platforms to understand and address these gaps to optimize user engagement and improve conversion rates. Moving forward, future work could focus on refining the user experience and addressing specific challenges identified in this analysis. Conducting more extensive user testing, especially on mobile devices, will be instrumental in uncovering issues and guiding targeted improvements. Implementing A/B testing on the checkout page may shed light on factors influencing the observed lower conversion rates, facilitating data-driven optimizations. This continuous refinement process aims to align the e-commerce platform with evolving user expectations and industry standards.

## 6. Conclusion

This analysis of user interactions and conversion metrics provides valuable insights for optimizing the e-commerce platform. The higher exit rates observed on mobile devices underscore the necessity for targeted improvements in mobile optimization, loading times, and user interface design. Addressing these concerns is aidful in ensuring a seamless user experience. Compatibility issues across different browsers, particularly on Safari and Firefox, require further investigation and implementation of necessary adjustments to enhance cross-browser performance.

The conversion rate analysis showed the varying effectiveness of different traffic mediums. Advertised sources (CPM and CPC) exhibit higher conversion rates, while referral and affiliate traffic present opportunities for improvement. Examining the user journey from product details to checkout revealed potential bottlenecks in the concluding stages of the conversion process. The identified disconnect between user interest and completed transactions signals the need for strategic enhancements in the checkout process, aiming to streamline and facilitate user transactions.

Our study underscores the significance of analyzing user engagement within e-commerce platforms to pinpoint potential gaps between initial interest and ultimate commitment. Recognizing these disconnects enables e-commerce platforms to devise strategies that enhance user engagement and improve conversion rates, fostering a more successful online shopping experience for consumers. The insights garnered from this analysis empower businesses to refine their platforms and align them more closely with user expectations, ultimately contributing to sustained growth and competitiveness in the digital marketplace.

**Figures**

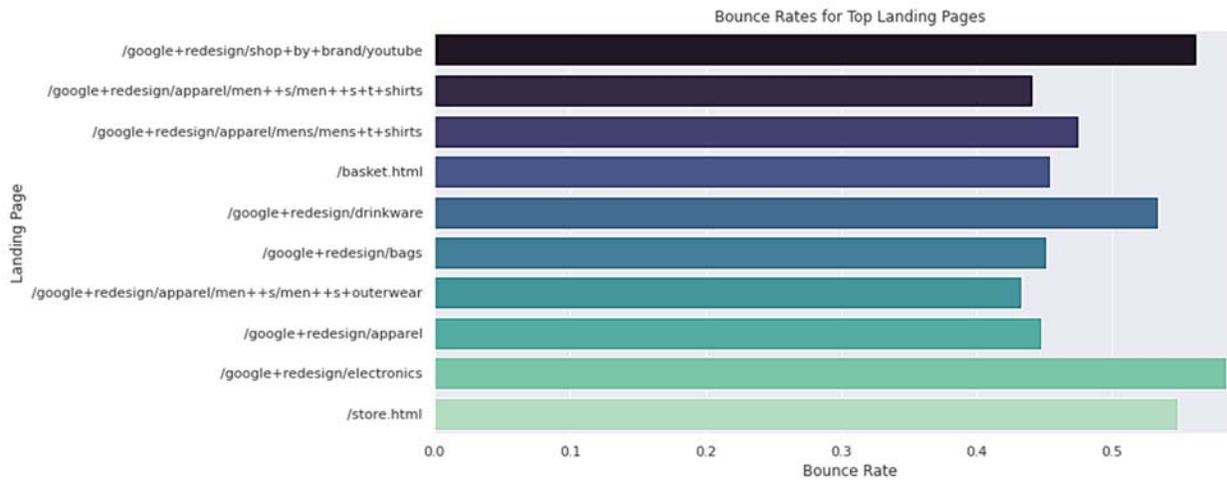

Fig. 1 Bounce Rate for Top Pages

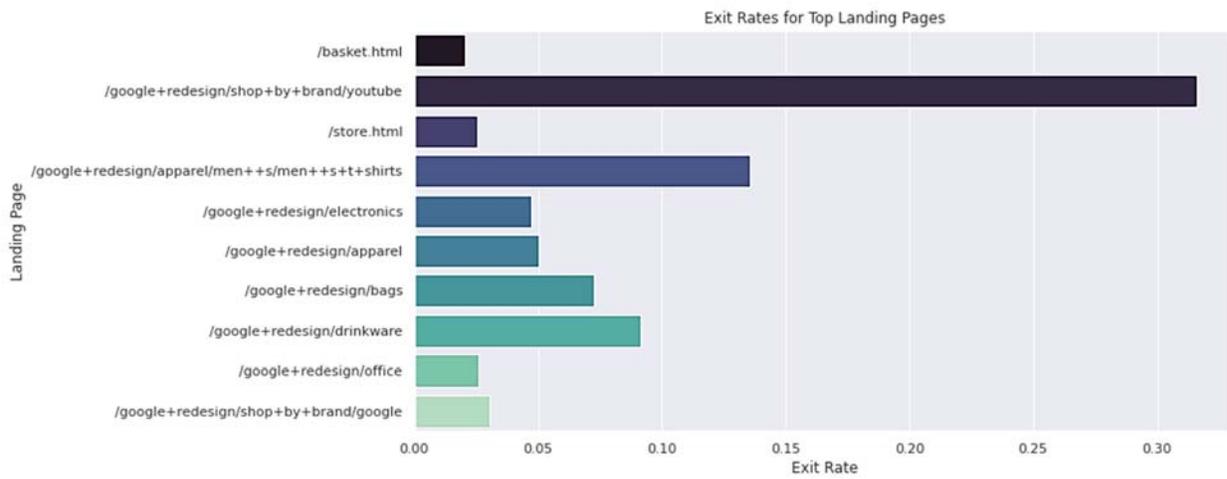

Fig. 2 Exit Rate for Top Pages

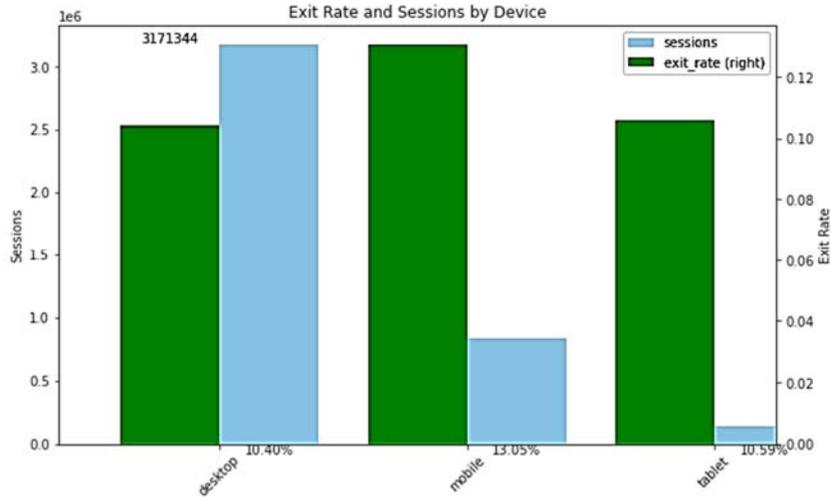

Fig. 3 Exit Rate and Sessions by Device

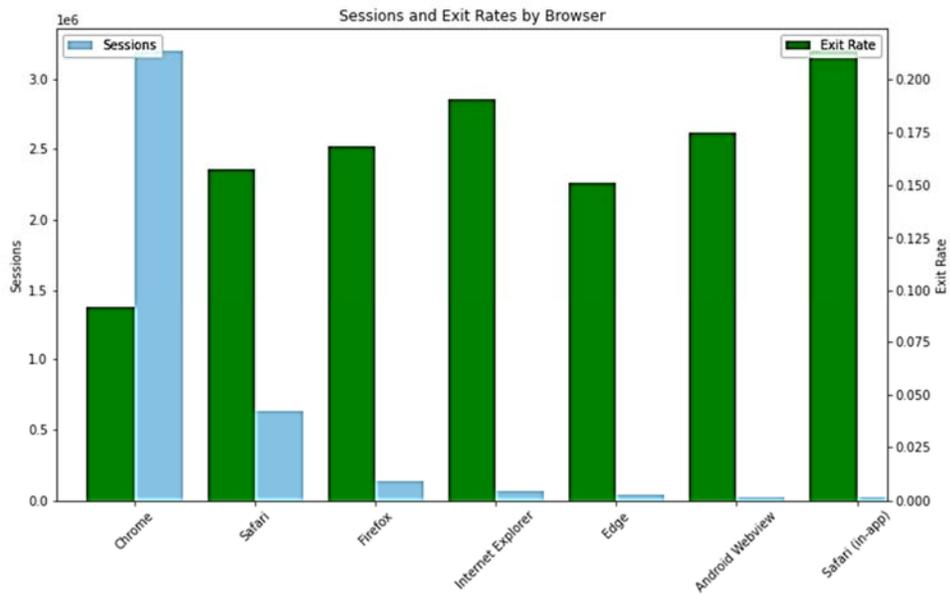

Fig. 4 Sessions and Exit Rate by Browser

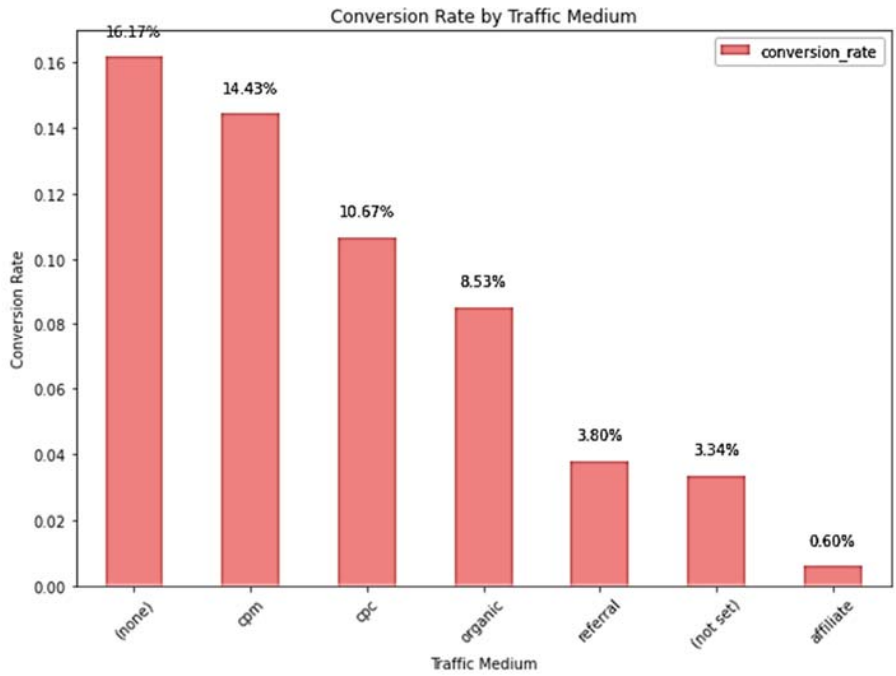

Fig. 5 Conversion Rate by Traffic Medium

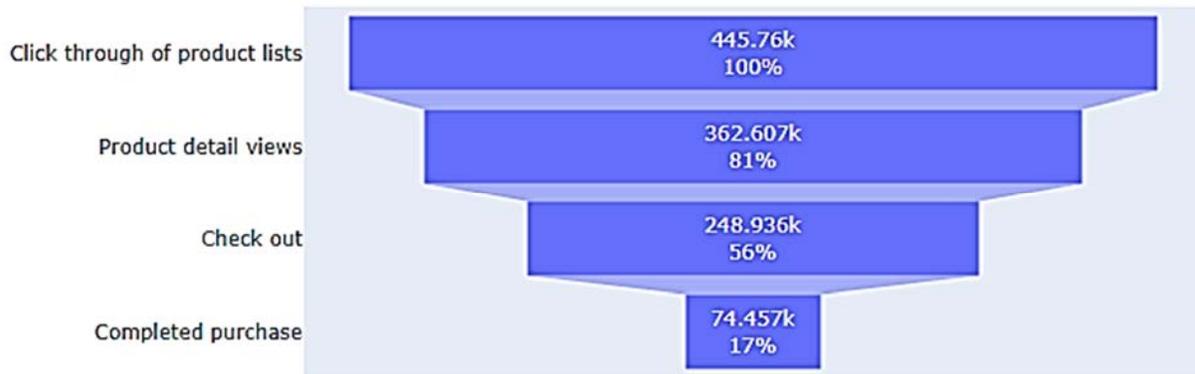

Fig. 6 Conversion Path